\documentclass[fleqn,usenatbib,useAMS]{mnras}

\usepackage{mathptmx}

\usepackage[T1]{fontenc}
\usepackage{ae,aecompl}

\usepackage{graphicx}	
\usepackage{amsmath}	
\usepackage{amssymb}	

\let\oldhref\href
\renewcommand{\href}[2]{\oldhref{#1}{\hbox{#2}}}

\newcommand{\p}{Pop~III }

\newcommand{\MO}{M_\odot}
\newcommand{\hMpc}{$ \, h^{-1} \rm Mpc$}

\newcommand{\hpc}{$ \, h^{-1} \rm pc$}

\title[IGM metal enrichment via Pop~III stars]{
Discrimination of heavy elements originating from Pop~III stars in $z = 3$ intergalactic medium
}
\author[T. Kirihara et al.]{
Takanobu Kirihara,$^{1, 2}$\thanks{E-mail: tkirihara@chiba-u.jp}
Kenji Hasegawa,$^{3}$
Masayuki Umemura,$^{4}$
Masao Mori$^{4,5}$
\newauthor
and Tomoaki Ishiyama$^{1}$\\
$^{1}$Institute of Management and Information Technologies, Chiba University, 1-33, Yayoi-cho, Inage-ku, Chiba 263-8522, Japan\\
$^{2}$Department of Physics, Graduate School of Science, Chiba University, Chiba 263-8522, Japan\\
$^{3}$Graduate School of Science, Nagoya University, Furo-cho, Chikusa-ku, Nagoya, Aichi 464-8602, Japan\\
$^{4}$Center for Computational Sciences, University of Tsukuba, Tennodai 1-1-1, Tsukuba, Ibaraki, 305-8577, Japan\\
$^{5}$Faculty of Pure and Applied Physics, University of Tsukuba, Tennodai 1-1-1, Tsukuba, Ibaraki, 305-8577, Japan\\
}

\date{Accepted XXX. Received YYY; in original form ZZZ}

\pubyear{2019}

\begin{document}
\label{firstpage}
\pagerange{\pageref{firstpage}--\pageref{lastpage}}
\maketitle

\begin{abstract}

We investigate the distribution of metals in the cosmological volume at $z\sim3$, in particular, provided by massive population III (Pop~III) stars using a cosmological $N$-body simulation in which a model of \p star formation is implemented. 
Owing to the simulation, we can choose minihaloes where \p star formation occurs at $z>10$ and obtain the spatial distribution of the metals at lower-redshifts. 
To evaluate the amount of heavy elements provided by \p stars, we consider metal yield of pair-instability or core-collapse supernovae (SNe) explosions of massive stars. 
By comparing our results to the Illustris-1 simulation, we find that heavy elements provided by \p stars often dominate those from galaxies in low density regions. 
The median value of the volume averaged metallicity is $Z\sim 10^{-4.5 - -2} Z_{\odot}$ at the regions.  
Spectroscopic observations with the next generation telescopes are expected to detect the metals imprinted on quasar spectra. 

\end{abstract}



\section{Introduction}

The first stars, so-called Population III (Pop~III) stars, are significantly important in galaxy formation and the cosmic reionisation history because they are the first luminous objects in the Universe. 
However, direct observation of their signal is still quite difficult \citep{Whalen2013}, since they end their lives in the early universe. 
If the low-mass end of the initial mass function is smaller than $0.8 M_{\odot}$, their lifetimes are longer than the cosmic age so that they should survive in the present-day Universe. 
Therefore, extensive surveys of field stars have been conducted to hunt low-mass metal-free or extremely metal-poor stars in the Milky Way \citep[][and the references therein]{Frebel2015}. 
Another way to approach the direct evidence of \p stars is the observation of the heavy elements ejected by the supernova explosions of massive \p stars. 
The \p originated metals would be undetectable in highly metal-enriched regions by Pop~I and Pop~II stars. 
On the other hand, \p originated metals should be detected in future observations if such metals are distributed in the regions less contaminated with the metals produced by Pop~I and Pop~II stars. 
In this study, we simulate the spatial distribution of the \p originated metals in $z=3$ intergalactic medium and examine the possibility to discriminate them from the heavy elements that are originated from Pop~I and Pop~II stars. 

Many theoretical studies have investigated the \p star formation \citep{Tegmark1997, Abel1998, Abel2002, Omukai1998, Omukai2000, Omukai2001, Omukai2003, Nakamura2001, Schneider2002, Bromm2002, Bromm2009, Yoshida2006, Yoshida2008, Hosokawa2011, Hosokawa2016, Stacy2012,Kuiper2018,Susa2019,Chon2019}. 
In the standard cold dark matter model, the formation of \p stars is expected to occur at $z\sim 20-30$ in dark matter minihaloes with $10^5-10^6 \, \MO$ \citep{Abel2002, Bromm2002, Bromm2009}. 
To investigate the nature of the first objects, many studies have simulated the formation of primordial clouds considering H$_2$ cooling. 
\p star formation in minihaloes lasts in reionisation epoch ($z\sim10$) due to the photodissociation of H$_2$ molecule. 
Though the mass function of \p stars is highly uncertain, some previous studies have provided the mass function of the massive side of \p stars using cosmological simulations \citep{Susa2014, Hirano2014, Hirano2015}. 
Very massive stars with $140\,\MO<M_{\rm \p}<260 \,\MO$ explode as pair-instability SNe (PISNe), and stars with $8\,\MO<M_{\rm \p}<40 \,\MO$ explode as Type-II core-collapse SNe (CCSNe) \citep[cf.][]{Chiaki2018}. 

Radiation hydrodynamic simulations of \p star formation in minihaloes have been conducted one by one adopting a zoom-in technique based on cosmological simulations \citep{Susa2014,Hirano2014, Hirano2015}. 
However, it is still challenging to perform cosmological hydrodynamic simulations for a cosmic time considering both \p star formation and evolution of galaxies or properties of IGM. 
\citet{Wise2012} conducted a challenging cosmological radiation hydrodynamic simulation considering \p and Pop~II star formation down to $z=7$ in 1 cMpc box. 
\citet{Chiaki2018} investigated the metal enrichment process of the environment from the formation of \p star-forming cloud to 1-50 Myr after an SN explosion embedding a massive \p star in a minihalo. 
In this study, we build a bridge between \p formation and IGM metal enrichment in a cosmological volume.

The spatial inhomogeneity of the IGM metals is significant but sill highly uncertain. 
A universal metallicity floor of $\sim10^{-3} Z_{\odot}$ at $z\sim3$ was suggested by using QSO spectra \citep{Songaila1997}, and the volume filling factor of metal enrichment of the IGM began to get attention \citep[e.g.][]{Madau2001,Mori2002}. 
However, the later observations did not detect such a floor at $z\sim2.5$ \citep{Simcoe2004}. 
They have reported that roughly 30 per cent of lines in the Ly$\alpha$ forest are [C, O/H] $\lesssim -3.5$. 
\citet{Dodorico2016} have investigated the metal abundance and distribution in the IGM at $z\sim2.8$ using a high S/N QSO spectrum and obtained association of \ion{C}{IV} absorption to \ion{H}{I} lines with their column density. 
They reported that all \ion{H}{I} lines in their sample with \ion{H}{I} column density with ${\rm log}\,N_{\rm HI}>14.8$ show  an associated \ion{C}{IV} absorption. 
43 per cent of \ion{H}{I} lines with its column density $14.0<{\rm log}\,N_{\rm HI}<14.8$ show  an associated \ion{C}{IV} absorption. 
It corresponds to the 40~kpc scale overdensity of $\sim$ 10.8 ($\gtrsim$2.1 for ${\rm log}\,N_{\rm HI}>14.0$) comparing with overdensity distribution of a cosmological hydrodynamic simulation. 
They also reported that metals with \ion{C}{IV} detection could lie at line-of-sight distance of $>300$~proper kpc from a star-forming galaxy. 
By measuring the correlations between the \ion{C}{IV}, \ion{O}{VI} and \ion{H}{I} absorption lines, the necessity of observation of void region was implied to explore \p metals \citep{Simcoe2004}. 
To understand IGM metal enrichment, further spectroscopic observations using such as the Prime Focus Spectrograph (PFS) \citep{Takada2014} and the Thirty-Meter Telescope (TMT) are required. 

In this study, we attempt to discriminate heavy elements originating from Pop~III supernovae in relatively low-redshift IGM, based on cosmological dark matter simulations. 
Although minihaloes harbouring Pop~III stars have been hard to resolve so far, a recent very high-resolution dark matter simulation allows us to identify minihaloes in which Pop~III stars form, and thereby explore the distribution of heavy elements as a result of Pop~III supernovae. 
Gas and metals distribution in the IGM blown away from galaxies via galactic wind has been investigated using cosmological hydrodynamic simulations \citep[e.g.,][]{Choi2011, Oppenheimer2012}. 
In these simulations, the strength of galactic winds driven by SNe feedback is controlled by some parameters such as the mass-loading factor from a star-forming region and the wind velocity. 
They showed that the distribution of gas and metals highly depends on the treatment of the feedback processes. 
Formation and evolution of galaxies have been investigated by large simulation projects such as the Illustris project \citep{Vogelsberger2014}, the EAGLE project \citep{Schaye2015} and the Illustris-TNG project \citep{Pillepich2018}, which implemented AGN (Active Galactic Nuclei) feedback by supermassive black holes at the centre of galaxies and also SNe feedback model. 
These simulations have achieved a certain degree of success in demonstrating galaxy evolution \citep[e.g.,][]{Torrey2013}. 
To compare the \p originated heavy elements with galactic metals, we analyse the results of the cosmological hydrodynamic simulation (Illustris-1). 
We provide mass abundance of heavy elements originated from \p stars and galaxies as a function of local density.

This paper begins with the introduction of numerical method, which contains a cosmological $N$-body simulation that resolves the minihaloes. 
To obtain the metal distribution originated in Pop~III stars, we establish the models of massive star formation that occur PISN or CCSN in the \p star-forming minihaloes (Section~2). 
In Section~3, we compare the spatial distribution of Pop~III originated metals with the results of the Illustris-1 cosmological hydrodynamic simulation at $z=3$. 
We discuss where we can discriminate the region that Pop~III metal dominates and the characteristic elemental abundance pattern in Section 4. 
We summarise our conclusions in Section 5. 

\section{Numerical methodology}

\subsection{Cosmological N-body simulation}
\label{sec:model}

We focus on metal enrichment of the IGM via supernova explosions of \p stars based on the hierarchical structure formation. 
For the purpose, we use a high-resolution cosmological $N$-body simulation conducted by \citet{Ishiyama2016} using a massively parallel TreePM code GreeM \citep{Ishiyama2009,Ishiyama2012}. 
The simulation adopts the cosmological parameter set of $\Omega_0=0.31$, $\Omega_{\rm b}=0.048$, $\lambda_0=0.69$, $h=0.68$, $n_s=0.96$ and $\sigma_8=0.83$ \citep{Planck2014}, and is run with $2048^3$ dark matter particles in a comoving box of $8$~\hMpc~on a side. 
The corresponding particle mass and gravitational softening length are respectively $5.13\times10^3\, h^{-1} \, \MO$, and $120$~\hpc. 
Dark matter haloes are identified by Friends-of-Friends (FoF) algorithm \citep{Davis1985} adopting a linking parameter of $b=0.2$. 
The details of constructing merger trees are described in \citet{Ishiyama2015}. 
The minimum halo mass consists of $32$ particles and consequently corresponds to $1.64\times10^5\, h^{-1} \, \MO$, which is good enough to resolve minihaloes where \p stars are expected to be formed.

\subsection{\p star formation}
\label{sec:popIIIgas}

To model \p star formation, we employ some \p formation criteria to be installed to our $N$-body simulation based on \citet{Ishiyama2016}.
Since \p stars can be formed when $\rm H_2$ cooling works well, we first set the lower threshold mass above which $\rm H_2$ molecules are sufficiently produced under the influence of the Lyman-Werner (LW) background radiation 
\citep{Machacek2001}. 
\citet{Ishiyama2016} assumed a spatially uniform and time-dependent flux $F_{\rm LW}(=4\pi J_{\rm LW})$ in unit of $10^{-21}\,{\rm erg}\,{\rm s}^{-1}\, {\rm cm}^{-2}\,{\rm Hz}^{-1}$ adopted 
\begin{eqnarray}
J_{\rm LW}=0.1\times 10^{1.8(-1-{\rm tanh}(0.1(z-40)))},
\end{eqnarray}
which is a fitting function derived to reproduce a cosmic reionisation simulation \citep{Ahn2012}. 
The lower threshold is defined by the critical virial temperature $T^{\rm crit}_{\rm vir}$: 
\begin{eqnarray}
\left(\frac{T^{\rm crit}_{\rm vir}}{\rm 1000 K} \right) =
0.36
\left[\left(\frac{F_{\rm LW}}{10^{-21}}\right) \,
(\Omega_b h^2)^{-1} \,
\left( \frac{1+z}{20} \right)^{3/2}
\right]^{0.22}. 
\end{eqnarray}
We exclude atomic cooling haloes and also exclude haloes merging with other haloes that have already experienced Pop~III star formation because such haloes are expected to be polluted by metals. 
In this study, we assume that zero or one massive \p star forms in each minihalo, and the star is represented by the most gravitationally bound particle in the halo. 
We obtain \p stars formed in minihaloes from $z=30$ until $z=10$. 

\begin{figure}
\includegraphics[width=\linewidth]{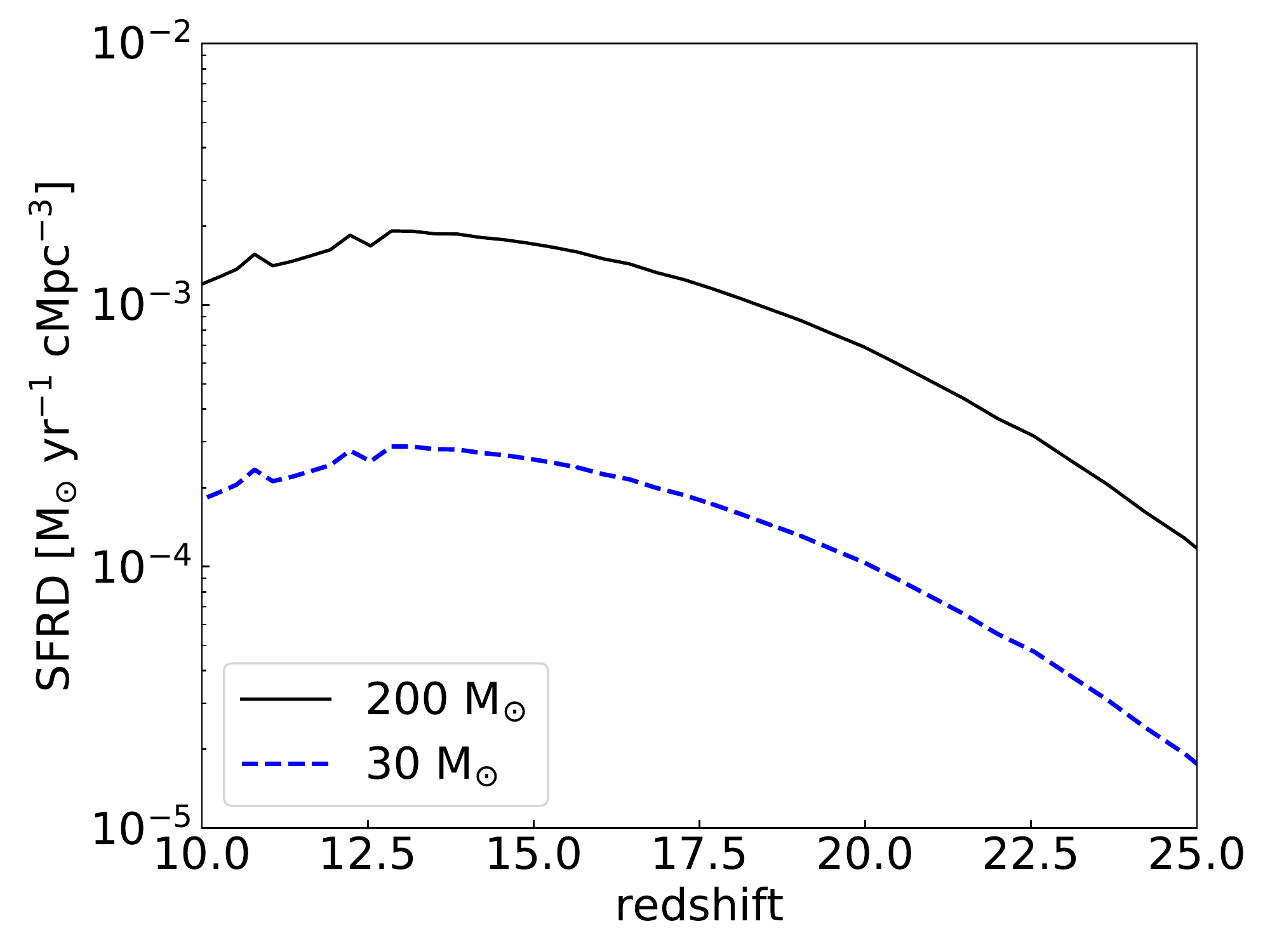}
\caption{Star formation rate density assuming a \p star instantaneously formed at the formation epoch of the parent minihalo. 
Black solid line and blue dashed line show the cases of \p stars with $200\, \MO$ and $30\, \MO$, respectively. 
}
\label{fig:SFRD}
\end{figure}

There are 3.8 million H$_2$ molecule cooling minihaloes formed in the cosmological simulation. 
Figure~\ref{fig:SFRD} shows redshift evolution of star formation rate density (SFRD) of \p stars. 
We here assume a \p star instantaneously formed at the collapse epoch of the parent minihalo. 
We show two cases that a massive star with $200\, \MO$ or $30\, \MO$ formed in each minihalo that satisfies the conditions of \p star formation. 
The SFRD gradually increases with decreasing redshift until $z=15$ and has a peak at $z\sim13$. 
The maximum SFRD is $\sim 10^{-3}$ for 200 $\, \MO$ case and several $10^{-4}$ for 30 $\, \MO$ case. 

Recent observations have shown that SFRD at $z=7-8$ is $\sim 10^{-2}\, {\rm \MO} \, {\rm yr}^{-1} {\rm Mpc}^{-3}$  \citep{Schenker2013, Madau2014, Ishigaki2018}. 
Thus it seems that SFRD in our model is not so different from the SFRD of star-forming galaxies. 
However, compared to recent state-of-the-art cosmological radiation hydrodynamics simulations \citep[e.g.,][]{Wise2012} in which SFRD at $z=10-16$ is $\sim 5\times 10^{-5} \, {\rm \MO} {\rm yr}^{-1} {\rm Mpc}^{-3}$, our SFRD for 200 $\rm \MO$ stars case is $1-2$ orders of magnitude higher than their SFRD. 
The difference is likely originated in our simplified LW background model that is independent of cosmic SFR. 
Neglecting the dependence of cosmic SFR could lead to the underestimate of LW background radiation field in the 200 $\rm \MO$ stars case. 
Moreover, such a high SFRD at $z=10-20$ may cause early reionisation which is conflict with the latest result by {\it Planck}. 
\citet{Madau2018} have indeed indicated that SFRD of $\sim 2 \times 10^{-3}$ ${\rm \MO} \, {\rm yr}^{-1} {\rm Mpc}^{-3}$ at $z\sim15$ overproduces ionising photons if the escape fraction of ionising photons is greater than 20 per cent \citep[see also][]{Madau1998}. 

We consequently adopt the following four models, including reduced SFRD models: 
\begin{itemize}
\item[(a)] \p candidates with 200 $\, \MO$, \par
\item[(b)] \p candidates with 200 $\, \MO$ whose SFRD are reduced by a factor of $0.1$, \par
\item[(c)] \p candidates with 30 $\, \MO$ and \par
\item[(d)] \p candidates with 30 $\, \MO$ whose SFRD are reduced by a factor of $0.1$. 
\end{itemize}
In the case of Model (a), mean escape fraction of ionising photons $\lesssim 20$ per cent is required. 

The yield masses of heavy elements from a $200$~$M_{\odot}$ star and a $30$~$M_{\odot}$ star are 113.7~$M_{\odot}$ and 6.7 $M_{\odot}$, respectively \citep{Nomoto2013}. 
In the case of Model (c), the yield metal is smaller by the factor of $\sim$ 0.06 than Model (a). 
In the Models (b) and (d), the SFRD of \p stars is artificially reduced to 10 per cent by randomly extracting 90 per cent of \p stars. 
\citet{Ishiyama2016} considered \p star formation under the homogeneous cosmic LW background radiation. 
However, it should be noted that star formation would be suppressed by nearby \p and Pop~II stars. 
Such a situation tends to occur in higher overdensity regions. 
Thus \p star formation rates in low-density regions might be underestimated if we randomly reduce \p stars. 

\subsection{Treatment of matter and metal distribution}

We mark particles that \p star formation occurs in high redshift ($z>10$) following the condition described in Section \ref{sec:popIIIgas}. 
The \p originated heavy elements are distributed at the positions of the marked particles at $z=3$. 
To compare heavy elements originaing from \p stars with those from galaxies at $z=3$, we refer to the results of Illustris-1 simulation, which implemented AGN feedback by supermassive black holes at the centre of galaxies and also SNe feedback model. 
The Illustris-1 simulation employed the moving-mesh code AREPO \citep{Springel2010} to follow the formation and evolution of galaxies \citep{Nelson2015}. 
It was run with $1820^3$ dark matter particles and $1820^3$ gas tracer particles in a comoving box of $75$~\hMpc. 
We coarse-grain the gas and metal distribution to $512^3$ mesh employing the spline kernel interpolation within the smoothing length. 
We adopt the SmoothingLength field provided in the snapshots of the public data. 

\begin{figure}
\includegraphics[width=\linewidth]{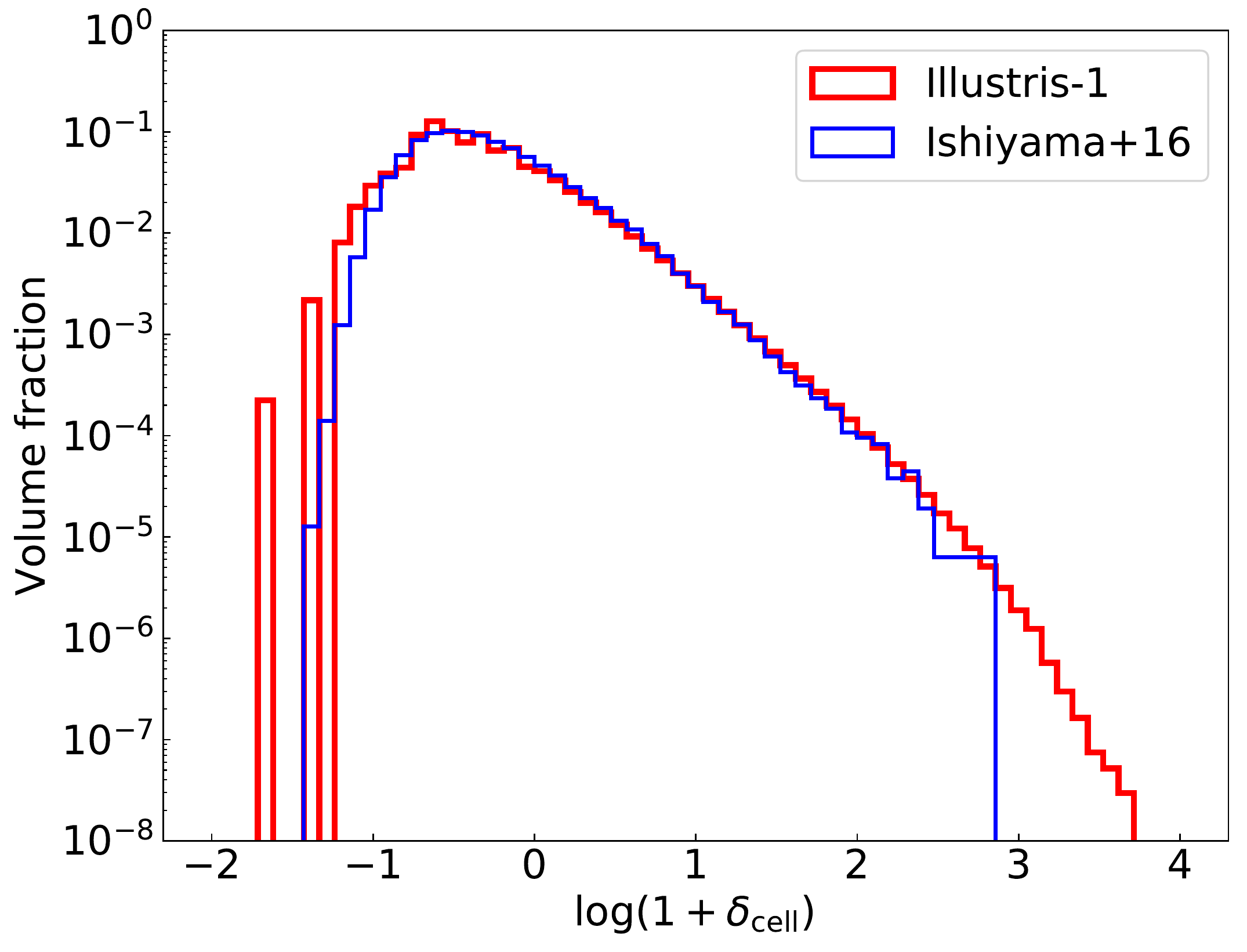}
\caption{
Volume fraction of local dark matter density divided into the coarse-grained mesh at $z=3$. 
The red line shows the result of Illustris-1 simulation \citep{Nelson2015}. 
The blue line shows the result of the high-resolution $N$-body simulation \citet{Ishiyama2016}. 
}
\label{fig:number_fraction}
\end{figure}

Figure~\ref{fig:number_fraction} shows the volume fraction of local dark matter density divided into the coarse-grained mesh. 
We define the local overdensity $\delta_{\rm cell}(\equiv \rho_{\rm cell, DM}/\overline{\rho_{\rm DM}}-1)$ as the dark matter overdensity of the coarse-grained mesh size ($\sim 50$~proper kpc) $\rho_{\rm cell, DM}$ to the cosmic mean dark matter density $\overline{\rho_{\rm DM}}$. 
For comparison, the distribution of $N$-body simulation is also coarse-grained to the same sized mesh.
The dark matter distribution of the $N$-body simulation and Illustris-1 simulation agree well each other over a wide density-range. 
The spike at log($1+\delta_{\rm cell}$)$\sim -1.6$ in the Illustris-1 simulation indicates that there is only one dark matter particle in the mesh. 
Since we focus on the lower-density region $-1< {\rm log}(1+\delta_{\rm cell})<2$, the difference at high-density region log($1+\delta_{\rm cell}$)$\gtrsim 2.5$ cannot affect our analysis. 
The difference would be originated in the rarity of haloes due to the difference of box size. 

\begin{figure}
\includegraphics[width=\linewidth]{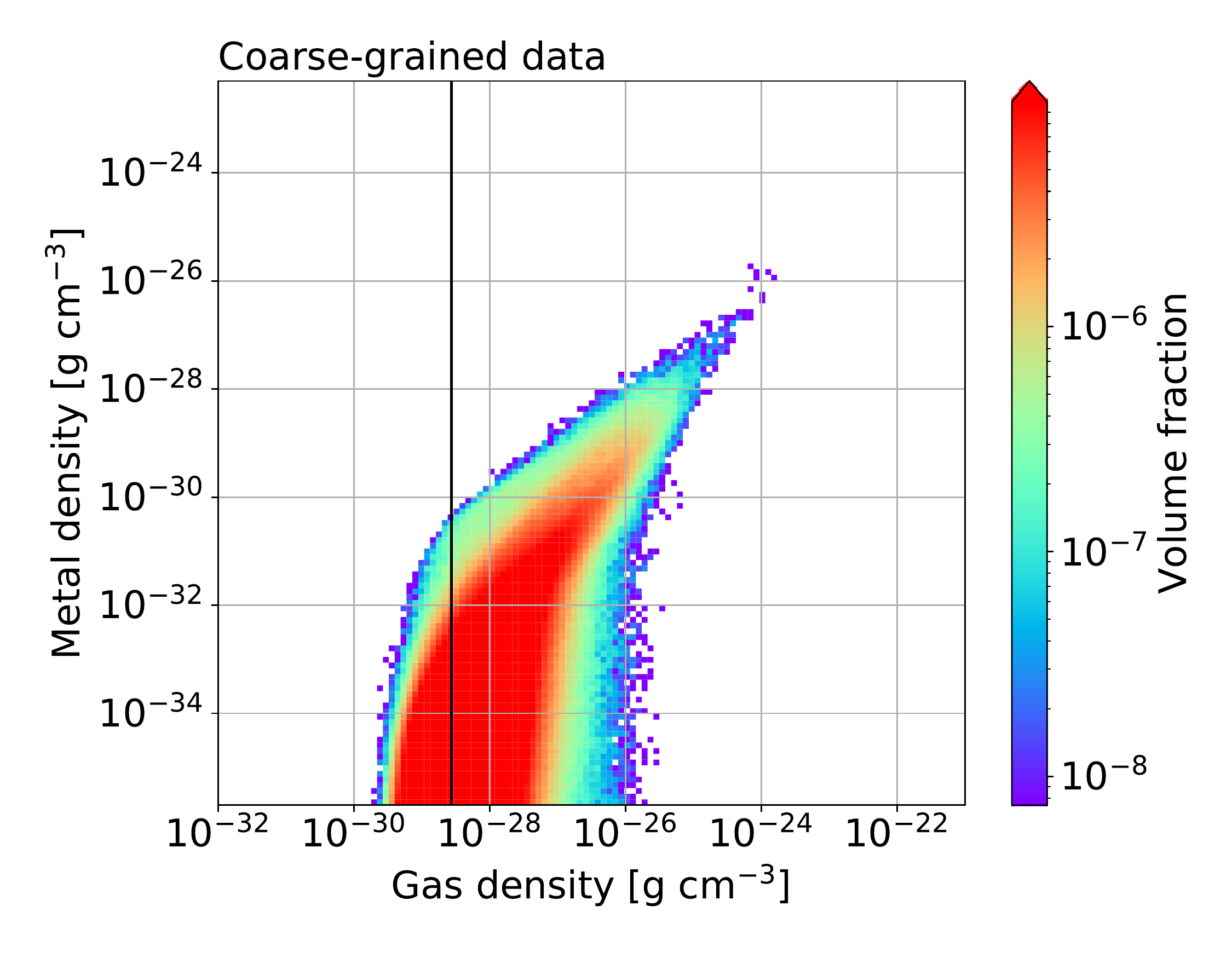}
\includegraphics[width=\linewidth]{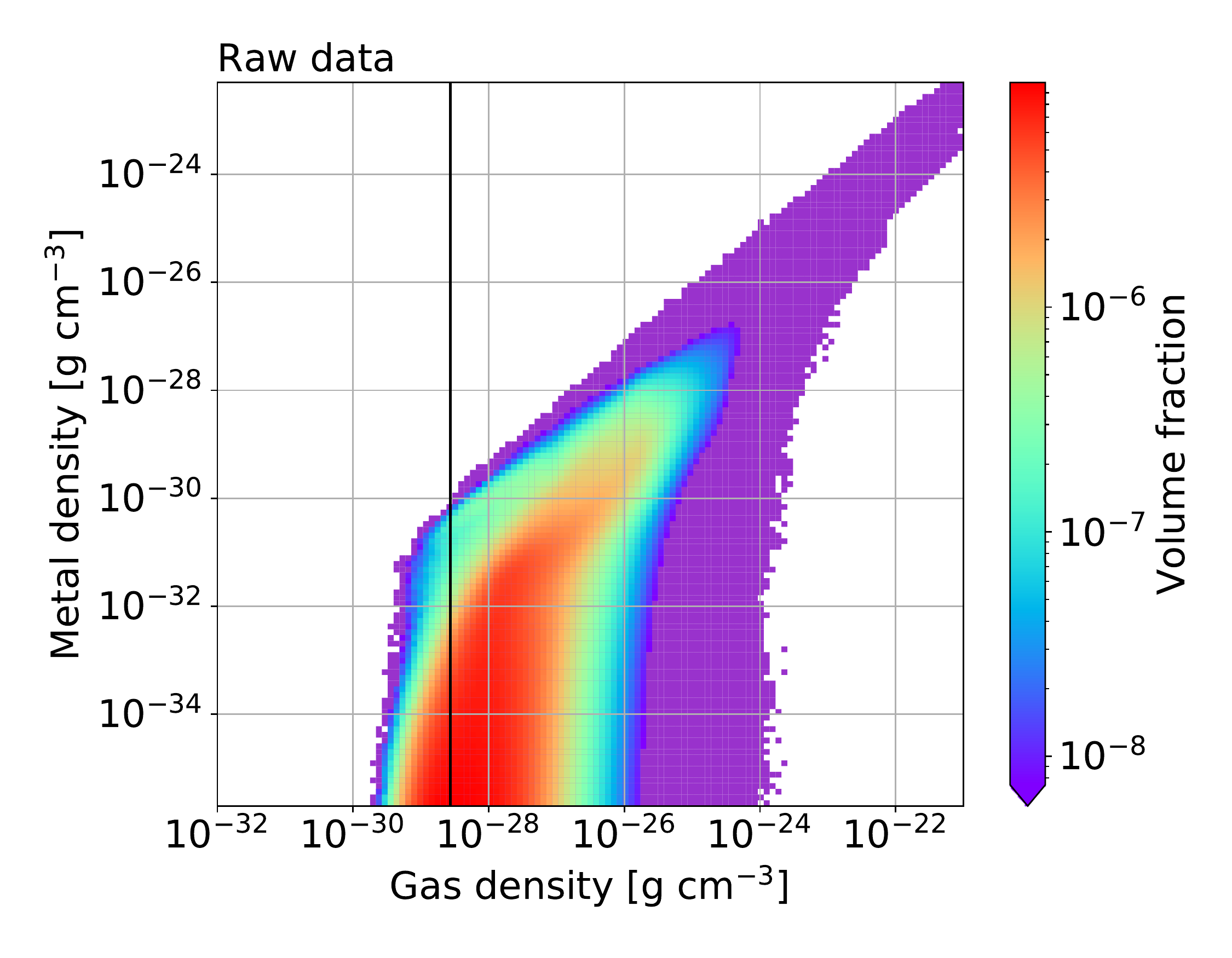}
\caption{
Volume fraction as functions of gas and metal density obtained from the Illustris-1 simulation. 
The top panel shows the physical quantities coarse-grained to $\sim 50$~proper kpc mesh, and the bottom panel shows the raw data of the moving mesh simulation. 
The vertical line indicates the cosmic mean baryonic density.  
}
\label{fig:IllustrisGasMetal}
\end{figure}

Fig.~\ref{fig:IllustrisGasMetal} shows volume fraction as functions of gas and metal mass density obtained by the Illustris-1 simulation. 
We compare between the original data of moving-mesh simulation and the coarse-grained data adopting the SPH-like treatment in post-process. 
As Figure~\ref{fig:IllustrisGasMetal} shows, the moving mesh simulation resolves the higher density region. 
The discrepancy at the higher density region is attributed to the coarse-grain approach that smoothes out the well-resolved structures. 
Due to the coarse-grained method, physical quantities in relatively higher density region get smaller. 
The vertical solid line in Figure~\ref{fig:IllustrisGasMetal} indicates the mean baryonic density. 
At lower gas density, the corresponding smoothing length is much larger than grid size. 
Therefore there is little difference between the physical quantity of Voronoi mesh and those of the coarse-grained data. 
Differences of volume fraction between the two models for $<10^{-34}{\rm g\, cm^{-3}}$ and $<10^{-33}{\rm g\, cm^{-3}}$ regions are 2.0 per cent and 0.77 per cent, respectively. 
Since we focus on the lower-density region, the difference at higher-density is not a serious problem. 

\section{Results}
\subsection{Metal enrichment}

\begin{figure}
\includegraphics[width=\linewidth]{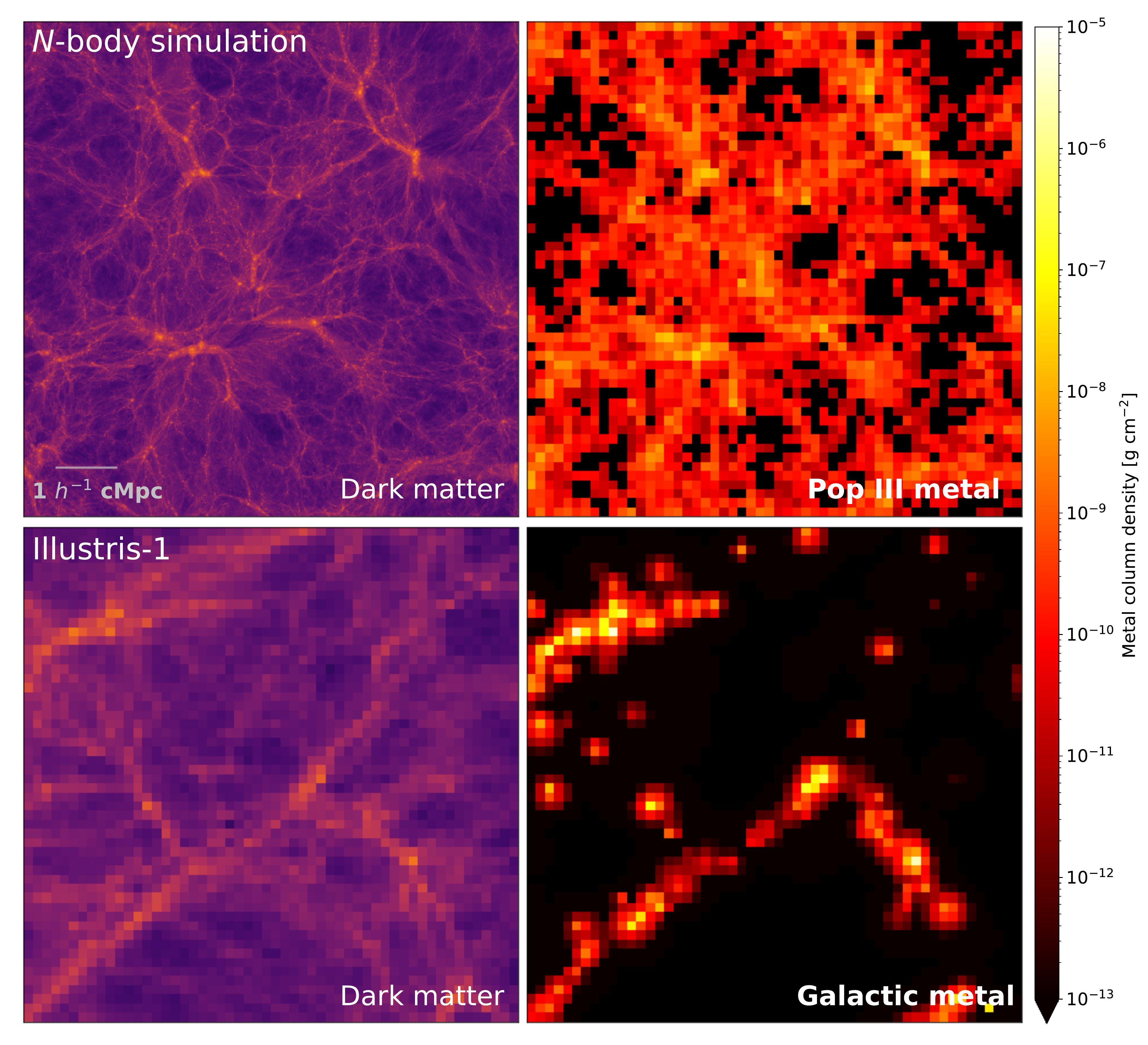}
\caption{
Top: Column density distribution of dark matter (left) and Pop~III originated metals (right) obtained by the cosmological $N$-body simulation at $z=3$. 
Bottom: Column density distribution of dark matter (left) and galactic metals (right) obtained by the Illustris-1 simulation. 
We use dark matter and metal distribution with a thickness of 160~proper kpc, and the side length is 8$ \, h^{-1} {\rm cMpc}$. 
}
\label{fig:map2}
\end{figure}

Figure~\ref{fig:map2} shows the column density distribution of dark matter and metals at $z=3$. 
We show the mass density distribution of heavy elements in the case of Model (a). 
The metal distribution is also divided into the same coarse-grained size, and the size of coarse-grained cells is larger than the radius of supernova shells of Pop~III stars. 
To illustrate the bottom panel of Figure~\ref{fig:map2}, we select an 8 $h^{-1}$ cMpc box, whose global overdensity is $\sim 0$, from the Illustris-1 simulation. 
Since the number density of minihaloes is high around galactic haloes, the column density of Pop~III originated metal is abundant there. 
The most striking result to emerge from the analysis is that Pop~III originated metals are also distributed very far from galactic haloes. 
It can be seen that heavy elements originating from \p stars are also distributed around the filamentary structure and also in the elaborate structures.
On the other hand, galactic metals are located by galactic haloes, and the distribution looks somewhat patchy. 
Pop~III star-formation takes place at small-scale density fluctuations at early epochs of the Universe ($z>10$). 
As a result of the nature of random Gaussian density fluctuations, minihaloes born in higher-$\sigma$ peaks are incorporated into massive haloes which constitute protoclusters of galaxies at $z=3$, whereas those born in lower-$\sigma$ peaks tend to be located in intergalactic space at $z = 3$. 
Metals originating from the latter minihaloes can be discriminated from metals in protoclusters. 
In other words, the merger history hardly proceeds to form massive haloes in intergalactic space. 
In this paper, we attempt to show where heavey elements orinating from \p stars dominate in the cosmic volume using the cosmological simulations.

\subsection{Comparison of heavy elements between \p origin and galactic origin}

\begin{figure}
\includegraphics[width=\linewidth]{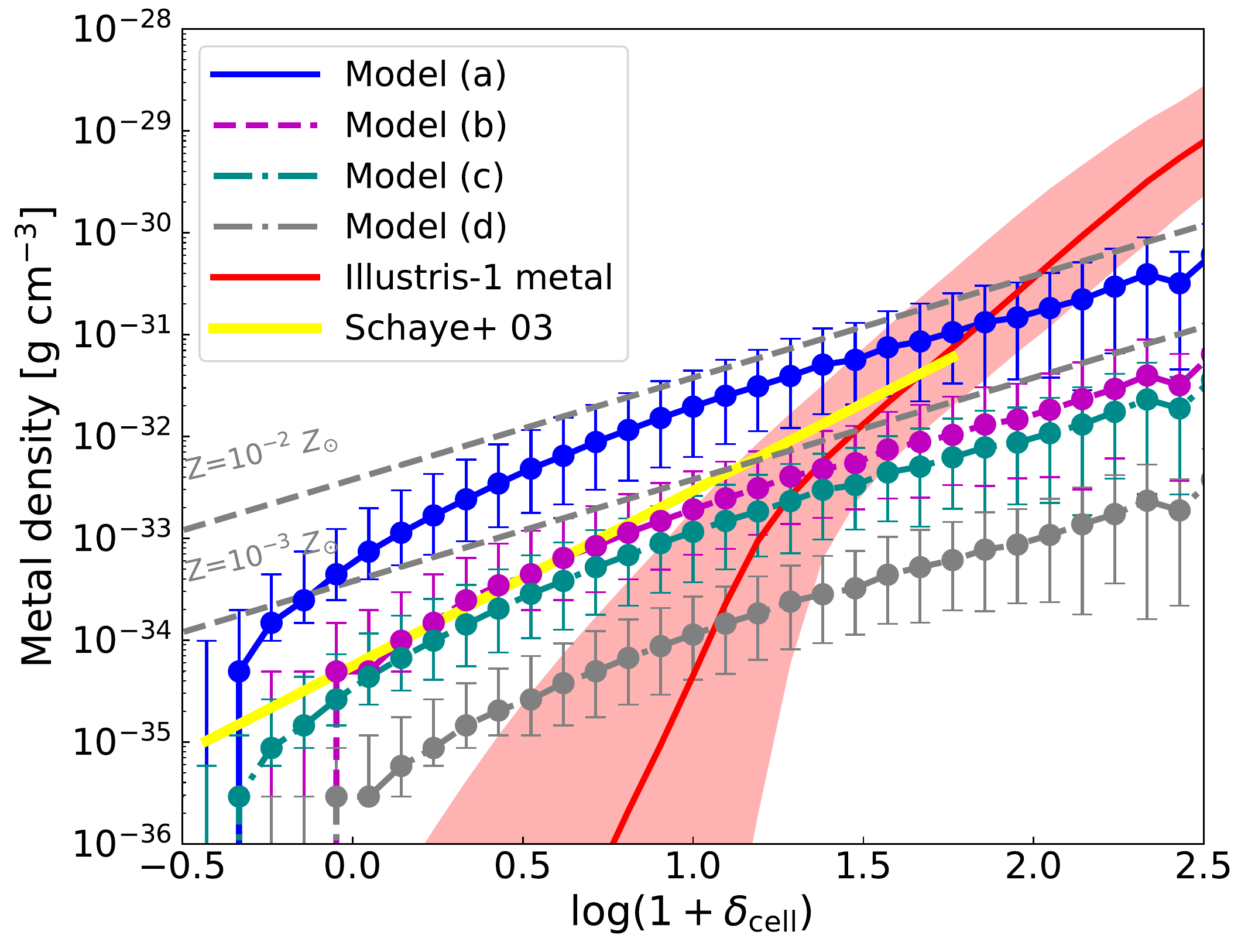}
\caption{
Metal density distribution as a function of local dark matter density. 
The overdensity $\delta_{\rm cell}$ is the dark matter density normalised by the cosmic mean dark matter density within the coarse-grained cell. 
The red shaded region shows the galactic metal distribution within 16 and 84 percentiles, and the solid line indicates the median distribution. 
The blue solid, magenta dashed, dark-cyan dash-dotted and grey dash-dotted lines correspond to Models (a), (b), (c) and (d), respectively. 
The errorbars shows the 16 and 84 percentiles of their distribution. 
The gray dashed lines indicate $10^{-2} Z_{\odot}$ and $10^{-3} Z_{\odot}$ assuming the gas to dark matter fraction of $ \Omega_{\rm b} / (\Omega_{\rm 0}-\Omega_{\rm b}) $ at the $\delta_{\rm cell}$. 
The yellow line shows the observational estimate of the overdensity-metallicity relation using Ly$\alpha$ forest \citep{Schaye2003}.   
}
\label{fig:gas_part_werr}
\end{figure}

Figure \ref{fig:gas_part_werr} shows the mass-density distribution of heavy elements as a function of $\delta_{\rm cell}$. 
If \p stars are formed as Model (a), the mass density of heavy elements originated in \p SN is a dominant contributor in ${\rm log}(1+\delta_{\rm cell})\lesssim 1.8$. 
In the cases of Model (b) and (c), \p originated metals dominate in ${\rm log}(1+\delta_{\rm cell})\lesssim 1.3$. 
All four \p models exceed the median distribution obtained by the Illustris-1 simulation in the low-density region. 

We overplot the overdensity-metallicity relation of the IGM for $2.5<z<3.5$ derived by \citet{Schaye2003}. 
Using the pixel optical depth technique, they measured the distribution of carbon as a function of overdensity $\delta$ in the range $-0.5<{\rm log}\,\delta<1.8$. 
They estimated the carbon metallicity [C/H]$=-3.47+0.72$(log $\delta -0.5$)
, which is calculated adopting the ionisation correction from \ion{H}{I} and \ion{C}{IV} and normalised by the solar abundance. 
To illustrate the observationally estimated relation, we assume that dark matter overdensity is the same as the baryonic overdensity and assume that the carbon metallicity matches the metallicity [Z/H]. 
It should be noted that the observed relationship has a similar positive correlation between the median metallicity of the IGM and overdensity to the relation derived by \p originated heavy elements. 
Metal abundance in Models (b) and (c) explains the observed relationship quantitatively in the lower density regions without fine-tuning. 

We roughly give the gas density of a metal-enriched region considering an independent SN ejecta. 
In practice, it is necessary to simulate the evolution of the SN remnant in the IGM environment with cosmic expansion. 
Here, we consider that the gas distribution in the low-density environment spread over 10~kpc owing to cosmic expansion and is assimilated with the IGM density in the zeroth-order approximation. 
We give an estimation assuming that we can detect a gas cloud with $ \sim 10^{-4}-10^{-3} Z_{\odot}$. 
For example, assuming the gas keeping the fraction $ \Omega_{\rm b} / (\Omega_{\rm 0}-\Omega_{\rm b}) $, $\rho_{\rm Z} = 10^{-34}\, {\rm g}\, {\rm cm^{-3}}$ corresponds to $ \sim 10^{-3.5} Z_{\odot}$ in an environment of $ \delta_{\rm cell}=0$. 

\begin{figure}
\includegraphics[width=\linewidth]{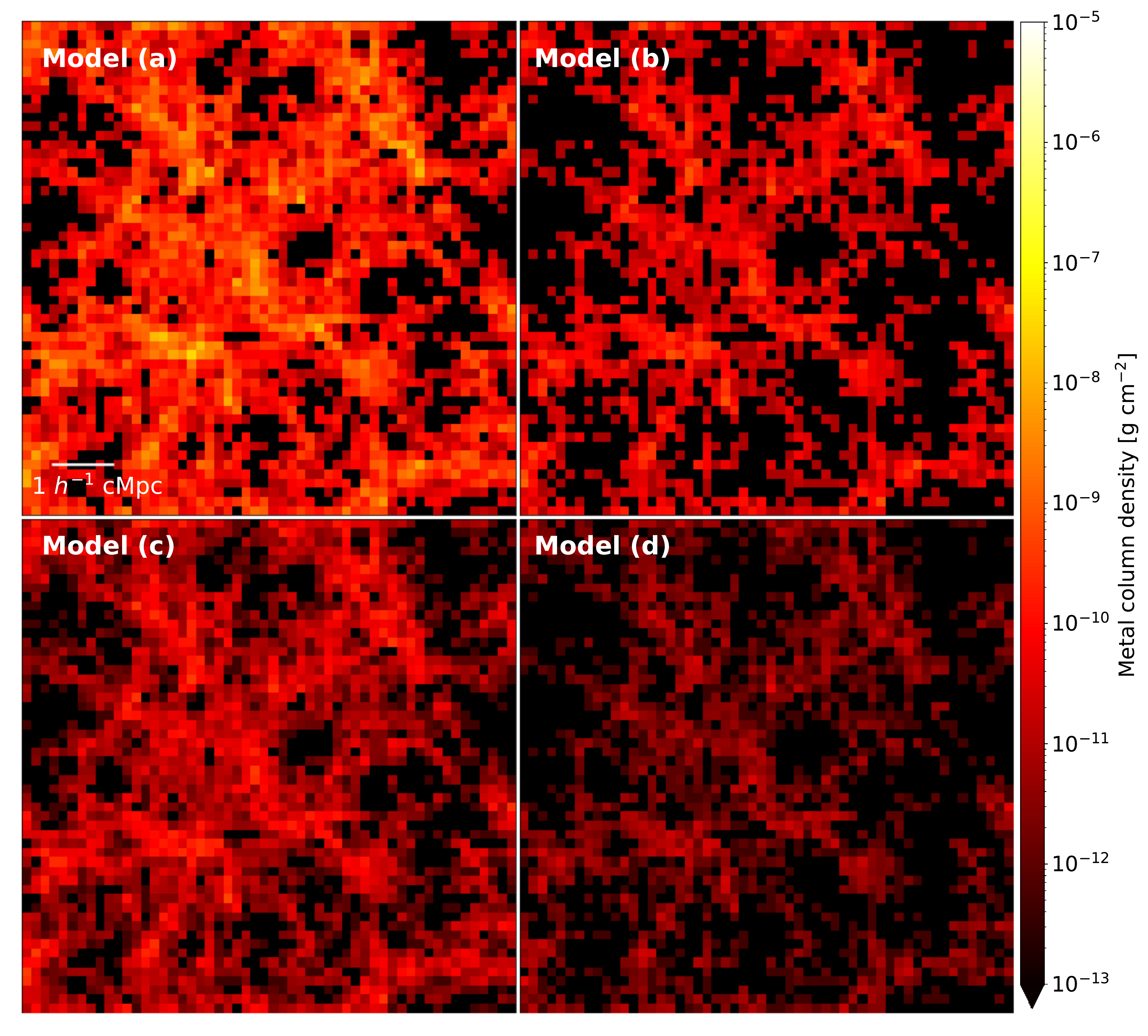}
\caption{
Same as top right panel of Figure~\ref{fig:map2}, but only the coloured cells are dominated by \p originated heavy elements. 
From top left to bottom right, the panels show the cases of Models (a), (b), (c) and (d), respectively. 
}
\label{fig:pop3dominated}
\end{figure}

Figure~\ref{fig:pop3dominated} shows column density distribution of the region that is dominated by \p originated metals. 
When illustrating this figure, we exclude the cell that heavy elements originating from galaxies exceed against \p origin heavy elements. 
We judge the excess only by the cell overdensity following the median lines of the overdensity-metal density relation as presented in Figure~\ref{fig:gas_part_werr}. 
We evaluate the mass density of heavy elements in each coarse-grained mesh and integrate density over a thickness of 160~proper kpc. 
This filtering removes the distribution of \p originated heavy elements in high $\delta_{\rm cell}$ region. 
Since the volume fraction of high $\delta_{\rm cell}$ regions is small (see Figure~\ref{fig:number_fraction}), there is little significant difference. 

In the cases of Models (b) and (d), the SFRD of \p stars is uniformly reduced by a factor of 0.1. 
Therefore, this treatment might excessively reduce \p star-formation in the low-density region. 
Nevertheless, the \p originated metals have an extended distribution that is unlike the patchy distribution associated with the galactic haloes as seen in the metal distribution of galaxies (see also the bottom panels of Figure~\ref{fig:map2}). 
In the sliced map, the integration of the average metal mass density of $10^{-34} $~g~cm$^{-3}$ over the thickness corresponds to the column density $\sim5\times 10^{-11}$ ~g~cm$^{-2}$. 

\begin{figure}
\includegraphics[width=\linewidth]{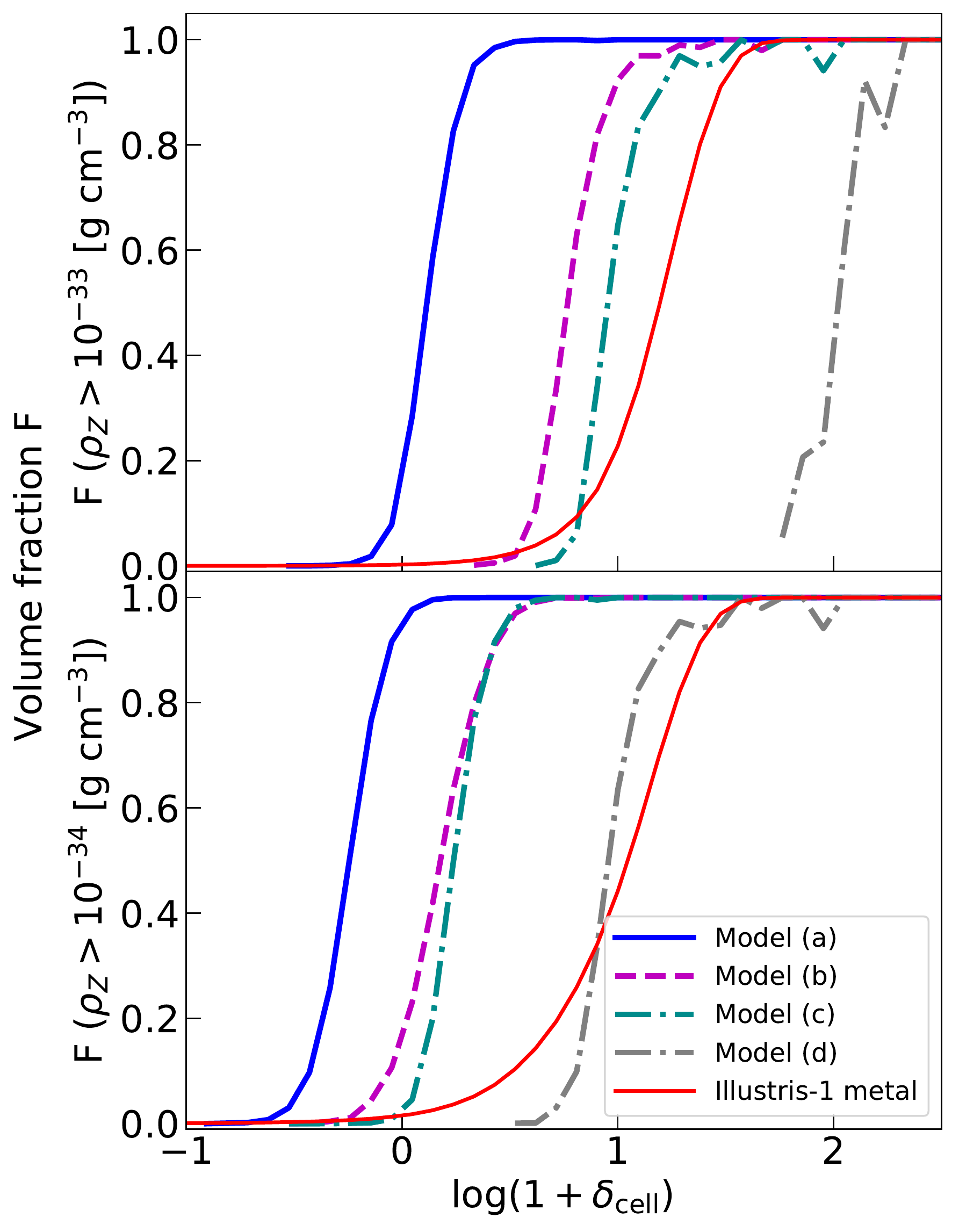}
\caption{ 
Volume fraction of region having the mass-density of heavy elements higher than $\rho_{\rm Z}>10^{-33} \, {\rm g}\, {\rm cm^{-3}}$ (Top panel) and $\rho_{\rm Z}>10^{-34} \, {\rm g}\, {\rm cm^{-3}}$ (Bottom panel). 
The red solid line shows the galactic metals obtained by the Illustris-1 simulation. 
The blue solid and dark-cyan dash-dotted lines are the cases of Models (a) and (c), respectively. 
The magenta dashed and grey dash-dotted lines show the SFRD reduced cases of Models (b) and (d), respectively. 
}
\label{fig:gas_metal_density_cover}
\end{figure}

It is a remarkable result that \p origin heavy elements have an enhancement at smaller $\delta_{\rm cell}$. 
Figure \ref{fig:gas_metal_density_cover} shows volume fraction $F$ of region having the metal-density higher than $\rho_{\rm Z}>10^{-33} \, {\rm g}\, {\rm cm^{-3}}$ (Top panel) and $\rho_{\rm Z}>10^{-34} \, {\rm g}\, {\rm cm^{-3}}$ (Bottom panel). 
Supposing the case that the metal density of $\rho_{\rm Z}>10^{-33} \, {\rm g}\, {\rm cm^{-3}}$ can be detected, heavy elements originating from \p stars in Model (a) exceed galactic metal in ${\rm log(1+\delta_{\rm cell})}\lesssim 1.8$. 
The volume fraction has $F>0.5$ in ${\rm log}(1+\delta_{\rm cell})\gtrsim 0$ region. 
The region ${\rm log(1+\delta_{\rm cell})}\sim 0$ occupies several per cent of whole volume (see Figure~\ref{fig:number_fraction}). 
Integrating volume with $0<{\rm log(1+\delta_{\rm cell})}< 1.8$, the volume corresponds to 15 per cent of the whole volume. 
In the cases of Models (b) and (c), volume fraction has $F>0.5$ in ${\rm log(1+\delta_{\rm cell})}\gtrsim 0.7$ region. 
The metal density of $\rho_{\rm Z}>10^{-33} \, {\rm g}\, {\rm cm^{-3}}$ is an indicator of the metallicity of $ \sim 10^{-3} Z_{\odot}$ if we adopt the gas mass fraction of $ \Omega_{\rm b} / (\Omega_{\rm 0}-\Omega_{\rm b})$ at ${\rm log(1+\delta_{\rm cell})}\sim 0.5$ region. 
If we can observe the region whose metal density is higher than $\rho_{\rm Z}>10^{-34} \, {\rm g}\, {\rm cm^{-3}}$, Pop~III origin heavy elements become more observable. 
Even if the observation reaches $\rho_Z > 10^{-34}$ g cm$^{-3}$, it is difficult in Model (d) to discriminate the Pop~III originated metals because they are mostly hidden in galactic metals.
We should observe a lower-density region with $\rho_Z <10^{-34}$ g cm$^{-3}$. 
Only around ${\rm log(1+\delta_{\rm cell})}\sim 1.2$ region has the volume fraction $F>0.5$ in the Pop~III metal dominated region. 

Considering the volume fraction distribution of the local density, \p metals would be observed easier at lower density region. 
It suggests that the low-metallicity absorption lines seen in the QSO spectrum, where the environment has a smaller $\delta_{\rm cell}$, may have originated from Pop~III stars.

\section{Discussion}
\subsection{Validity of numerical approach}

Overdensities at the absorbing objects are estimated by converting from \ion{H}{I} column density in QSO absorption spectrum \citep[e.g.][]{Schaye2003}. 
It should be noted that the overdensity and metallicity in our plots are calculated by coarse-grained physical quantities. 
To determine the observational metallicity, a gas distribution which is perhaps independent of $\delta_{\rm cell}$ is required. 
Therefore, when discussing metallicity, we may have to model the clumpy absorbers in a certain density in the environment of $\delta_{\rm cell}$.
More than one clouds are possibly located in a cell, and they would be localised to a certain extent. 
As far as we focus on global metal distribution with the resolution of the coarse-grained size, they can not be severe problems. 

We modelled metal enrichment of IGM caused by SN explosions of massive \p stars. 
Such a SN explosion occurs in a dark matter minihalo. 
Considering the gravitational binding energy of $10^5\, \MO$ halo, gas within the halo is easily blown away from the halo when a PISN explosion occurs \citep[e.g.,][]{Jaacks2018}. 
\citet{Chiaki2018} investigated the enrichment process from the formation of \p star-forming cloud to 1-50 Myr after a SN explosion embedding a massive \p star in a minihalo. 
They showed that if a PISN occurs in a minihalo with $3\times 10^5 \,\MO$, the yield metal can easily reach $\sim 1$~kpc at $50$~Myr after the SN. 
\citet{Wise2012} conducted a radiation hydrodynamics simulation of the formation of \p stars and Pop~II stars down to $z=7$ using an adaptive mesh refinement scheme in a 1~Mpc box. 
They obtained that one PISN with the stellar mass of $100 \, \MO$ enriches the halo and surrounding $\sim 5$~kpc to a metallicity of $10^{-3}$ ${\rm Z_{\odot}}$ as of $z=7$. 
If a CCSN with the stellar mass of $30 \, \MO$ occurs in a relatively massive minihalo, a part of the ejected gas and metals may fall back to the minihalo \citep[see][]{Chiaki2018}. 
If we consider the expanding radius of metals, the realistic distribution might be more localised within the mesh or might straddle cells. 
Such effects should be confirmed after updating the feedback process of \p stars in the environment. 

Implementation of the galactic wind modelling largely affects the metallicity evolution of the IGM \citep{Choi2011,Oppenheimer2012}.
Recent large projects of cosmological hydrodynamic simulations such as the Illustris and the EAGLE reproduced some representative observed properties of low-redshift galaxies (e.g., Tully-Fisher relation \
\cite{Tully1977}).
In the Illustris simulation, they implemented feedback from star formation kinetically so-called ``momentum driven'' wind.
The wind velocity was scaled to the velocity dispersion of the dark matter. 
They emphasised that it required very strong SN and AGN feedback to reproduce stellar mass-halo mass relation \citep{Vogelsberger2013}.
On the other hand, in the EAGLE simulation, the feedback was implemented thermally rather than kinetically, and the feedback energy was varied with local gas properties.
However, the column density of \ion{C}{IV} and \ion{O}{VI} obtained by the simulations is smaller than observed data at lower redshift, although they dought some systematics in observations.
To transport more metals from galaxies into the IGM, they imply the necessity of more effective feedback \citep{Schaye2015}. 
We consider the results of the Illurstris simulation in which galactic metals are strongly blown off from the galaxies. 

\subsection{Detectability in high-z QSO spectrum}

Absorption lines in QSO spectra are fundamental to understand the physical properties of the IGM. 
Owing to the Sloan Digital Sky Survey, the observed number of QSOs is increasing \citep{Fan2006,Richards2009,Paris2018}. 
\citet{Paris2018} reported 526 356 QSOs detected over 9376 deg$^2$. 
Approximate 5$\times 10^4$ QSOs have been discovered in $z>3$, and they are potential objects to be used as the background sources. 
We attempt to estimate how often ray from the QSO passes through the \p dominated region when we ideally assume that spectrum data with sufficient S/N is obtained. 
To calculate the volume that Pop~III originated heavy elements dominate, we integrate the volume fraction F multiplying the $\delta_{\rm cell}$ distribution shown in Figure~\ref{fig:number_fraction}. 
If we assume that $\rho_{\rm Z}>10^{-34}$ g cm$^{-3}$ metals can be detected, the \p metals dominated volume fraction is $3.2\times 10^{-1}$ in Model (a). 
In the case of Models (b), (c) and (d), the volume fractions are $1.1\times 10^{-1}$, $8.0\times 10^{-2}$ and $1.7\times 10^{-3}$, respectively. 
We give the coarse-grained volume to the $2/3$ as the cross-section of coarse-grained mesh to the background source. 
The impact factor is 18 in Model (a) in the line-of-sight with $\Delta z=0.012$ at $z=3$ per QSO spectrum. 
The numbers of Models (b), (c) and (d) are 5.8, 4.4 and 0.093, respectively.  

In this estimation, we ignore the effect of multiple absorption lines within the coarse-grained mesh. 
Further studies are required to estimate the feasibility considering the ionising state of heavy elements provided by hydrodynamic simulations.

\subsection{Insights from elemental abundance pattern}

\begin{figure}
  \includegraphics[width=\linewidth]{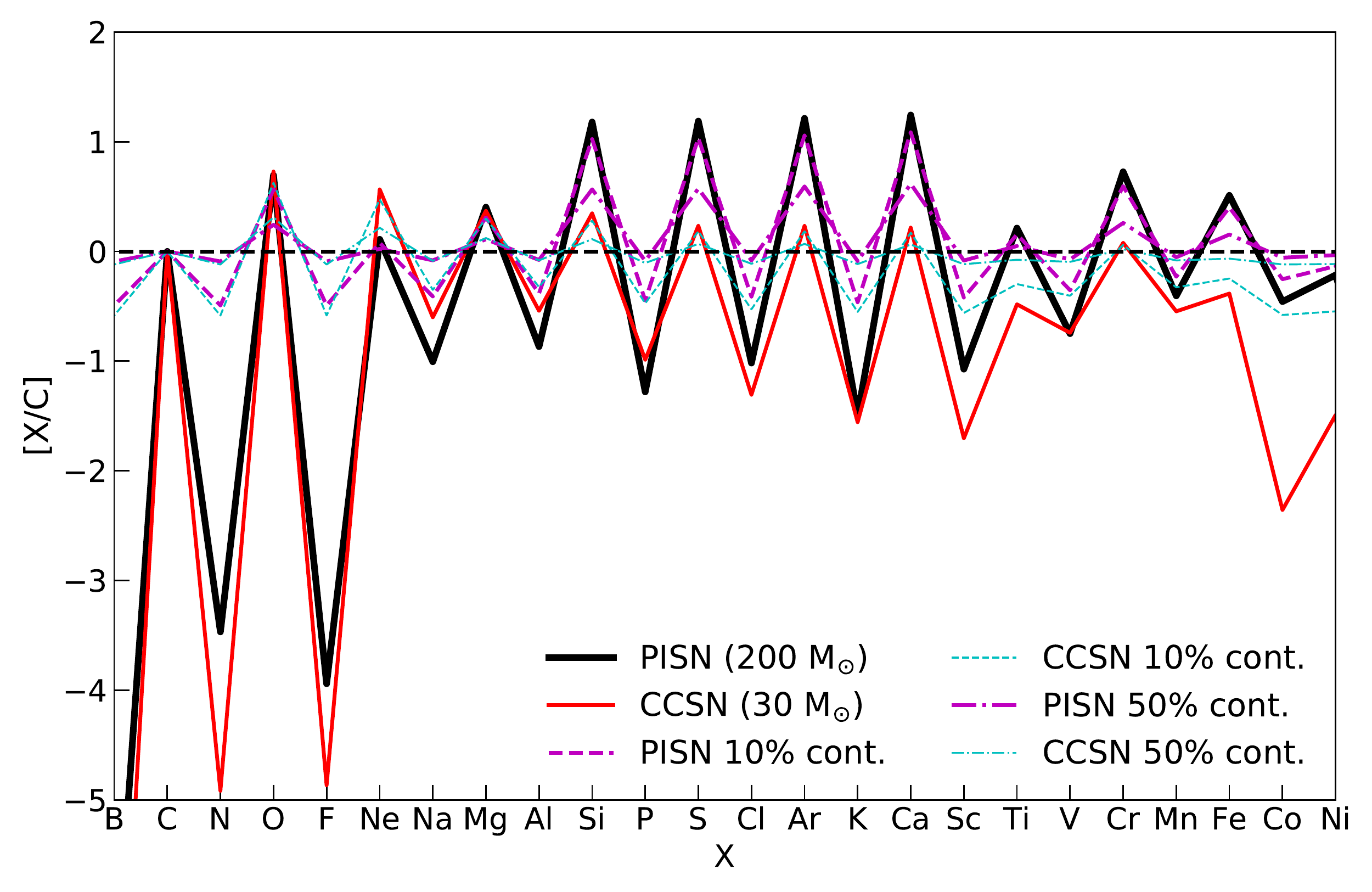}
\caption{
Elemental abundance patterns of PISN (200$\, \MO$) and CCSN (30$\, \MO$). 
Each elemental abundance is normalised by C abundance. 
The dashed lines show the elemental abundance patterns of Pop~III SNe yields contaminated by the galactic metals with the mass fraction of 10 per cent.  
As a reference model, we also plot the 50 per cent contaminated abundance pattern with dash-dotted lines. 
}
\label{fig:abundancepattern}
\end{figure}

An important point is whether we can notice that the observed heavy elements are not originated in galactic outflow when we observe Pop~III metal dominated regions. 
To confirm whether the heavy elements originated from Pop~III yields, it is necessary to obtain several elemental abundances when the metal distribution can be obtained in the future via absorption lines. 
This is because we cannot know whether the corresponding absorption is originated in Pop~III yields from the metallicity alone, and we need to look at the elemental abundance pattern. 
If ideally all the elemental abundance below iron can be obtained, Figure~\ref{fig:abundancepattern} implies the difference between PISN-, CCSN- and galactic-originated elemental abundance pattern. 
The original yield table to plot these elemental abundance patterns of PISN (200$\, \MO$) and CCSN (30$\, \MO$) is provided by \citet{Nomoto2013}. 
The abundance ratio is defined as [X/C] = ${\rm log}(n_X/n_C) - {\rm log}(n_X/n_C)_{\rm \odot}$, where $(n_X/n_C)$ and $(n_X/n_C)_{\rm \odot}$ are the number fraction of an element X relative to carbon of Pop~III models and solar abundance model, respectively. 
For example, the abundances of odd elements such as N produced by a PISN or a CCSN are much lower than those of the solar abundance pattern. 
We assume that the elemental abundance pattern of the galactic metals is the solar abundance pattern \citep{Grevesse1998,Allende_Prieto2001,Allende_Prieto2002,Holweger2001}. 
When each abundance is normalised by carbon abundance, even elements (e.g., O, Mg, Si, S, Ar and Ca) of a PISN (a CCSN) are greater by a factor of 3-20 (2-5) than those of the galactic-originated heavy elements. 
Absorption lines of C, O and Mg abundances are often targeted to investigate IGM. 
\ion{Si}{II} $ \lambda $1260 is detected in z $ \sim $3 damped Ly$\alpha$ (DLA) system \citep{Penprase2010,Cooke2011}. 
Further observations of heavy elements in absorbers such as Ly$\alpha$ forest at lower-density regions are required. 
The abundance ratio between C, N and O is useful to confirm at least not from galactic origin heavy elements.

One of the most significant points is whether Pop~I and Pop~II alone cannot explain the obtained metal abundance. 
If we look at the distribution of the median mass density of heavy elements in Models (b) and (c), the Pop~III originated heavy elements are about an order of magnitude larger than galactic originated metal in ${\rm log}(1+\delta_{\rm cell})\sim 1$. 
Heavy elements originating from galaxies might slightly contaminate the observed metal. 
We show the elemental abundance pattern of the region 10 per cent and 50 per cent contaminated by the galactic metals. 
As Figure~\ref{fig:abundancepattern} shows, we can distinguish them using the abundance ratio between C, N and O if the \p originated metal contributes more than the half of the mass of heavy elements in the region. 
It has been claimed that the abundance ratios between C, N and O are useful to recognise Pop~III originated heavy elements \citet{Ryan1996, Cayrel2004}. 
We explore the possibility of the discrimination of Pop~III originating heavy elements against the contamination of galactic metals by analysing the spatial distribution of metals obtained by numerical simulations. 
We again indicate that the elemental abundance ratios are significant tracers to explore the relics of Pop~III stars.
If N abundance is detected, the gas would be contaminated by galactic metal. 
If Pop~III stars are rotating, a stronger internal mixing boosts N production via CNO cycle in the H-burning shell, which changes the abundance ratios between C, N and O \citep[e.g.][]{Meynet2006,Ekstrom2008,Heger2010}. 
Stellar rotation of Pop~III star has recently begun to be investigated using high-resolution magnetohydrodynamic simulations of angular momentum transport between the primordial gas disc and the stellar rotation \citep{Hirano2018}. 
Besides, if PISN-originated heavy elements are dominated, the gas would be Si-rich, S-rich, Ar-rich, Ca-rich, Cr-rich and Fe-rich even in the contaminated cases. 
We will discuss the feasibility of the detecting absorption lines solving the physical state of gas and heavy elements by hydrodynamic simulation in our future work. 

If the $\delta$ on the 8~$h^{-1}$ cMpc scale, which corresponds to the boxsize of our simulation, is a little smaller, \p originated metals may be more dominant so that we should investigate the effect in the future research. 
Stacking of gas might be required to detect metal lines in such a lower density region. 
It can be a valid window to observe such a region in IGM that elemental abundances of S and Si are rich compared with surrounding metals. 

\section{Summary}

In this paper, we have investigated the relics of \p stars in the IGM at $z=3$. 
We examine metal enrichment of IGM via \p supernovae based on the hierarchical structure formation of the Universe employing a cosmological $N$-body simulation. 
It is the great advantage of using the high resolution $N$-body simulation that we can follow \p star-forming minihaloes and obtain the spatial distribution of \p originated metals. 
We treat the star formation rates of Pop~III stars as to be roughly consistent with the previous hydrodynamic simulations that proposed their initial mass function.
To see the difference in the results of CCSN and PISN, we embed single stellar mass (30 $M_{\odot}$ and 200 $M_{\odot}$).
As a result, the \p originated metals are distributed not only within galactic haloes but low-density regions that are located at large distances from galactic haloes. 
To compare the distribution of \p originated metal with galactic metal quantitatively, we analysed the results of Illustris-1 simulation.
In Models (b) and (c), \p originated heavy elements dominate in a region with $0\lesssim {\rm log }(1+\delta_{\rm cell})\lesssim 1.3$. 
The corresponding median metal density is $10^{-34} \, {\rm g}\, {\rm cm^{-3}}\lesssim \rho_{\rm Z}<10^{-33} \, {\rm g}\, {\rm cm^{-3}}$. 
The metallicity of such region is $ \sim 10^{-3.5} Z_{\odot}$ if we adopt the gas mass fraction of $ \Omega_{\rm b} / (\Omega_{\rm 0}-\Omega_{\rm b}) $ in its local overdensity. 

Once star-formation occurs, UV radiation in the Lyman-Werner bands from the stars would suppress \p formation in surrounding minihaloes. 
Such a situation tends to occur in relatively active star-forming regions. 
We reduce the effective star formation rate, but the modelling does not depend on the environment. 
Further studies focusing on the feedback process in the various environment are required to update this point. 
The number of Pop~III star-forming minihaloes achieve numerical convergence as long as we run with current models that are consistent with \citet{Hirano2015}.
A similar convergence test has been conducted using hydrodynamic simulations under different setup \citep{Schauer2019}.
Multiple Pop~III star-formation in each minihalo via fragmentation of circumstellar disc \citep[e.g.][]{Turk2009, Greif2012, Susa2019} can reduce the effective number of very massive stars. 
It is required that statistically reliable IMF information is examined using high-resolution cosmological hydrodynamic simulations of Pop~III stars. 
Observations of gravitational-wave events might limit the fraction of multiple Pop~III star-formation in each minihalo \citep{Kinugawa2014,Tagawa2015}. 

Considering the effect of cosmic expansion for the IGM that the ejecta sweeps up, the yield metals are gone off to the outward of minihaloes. 
The environmental density of IGM would affect the propagation \citep[e.g.,][]{Ikeuchi1983}. 
We analyse the smoothed distribution of metals to the mesh-size ($\sim 50$~proper kpc), therefore, the contribution of expanding radius of each shell to the global distribution of metals is small. 
On the other hand, if a \p hosting minihalo isolates until low-redshift, the gas and metals are expected to be observed as a localised absorber. 
In order to estimate the accurate feasibility of detecting the metals taking in such effects, further works of high spatial and time resolution cosmological radiation hydrodynamic simulations are required. 

We provide metal distribution originated in \p stars with 200 $\MO$ or 30 $\MO$ stars. 
This approach will be useful in expanding our understanding of how the metals are distributed in the cosmic volume. 
Although it has not been concluded that which mass stars are typical or there exist several typical masses \citep[e.g.,][]{Hirano2015}, 
we can obtain the region with a higher mass density of heavy elements than galactic metals in all our modellings. 
It is essential whether we can distinguish the \p yield heavy elements and galactic metals when we observe \p originated metal dominated regions. 
We take the elemental abundance pattern of SNe into account and confirm that the observed metals should have a characteristic elemental abundance pattern, which is not originated in galactic metals, even for partially contaminated cases.

\section*{Acknowledgements}

We thank Kazuhiro Shimasaku, Masami Ouchi, Nao Suzuki, Tohru Nagao, Akio Inoue and Gen Chiaki for helpful comments and discussions. 
We also thank an anonymous referee who provided valuable comments that improved the paper. 
Numerical calculations were partially carried out on Aterui supercomputer at Center for Computational Astrophysics, CfCA, of National Astronomical Observatory of Japan and the K computer at the Riken Institute for Computational Science (Proposal numbers hp150226, hp160212, hp170231, hp180180 and hp190161).
This research has been supported in part by MEXT as "Priority Issue on Post-K computer" (Elucidation of the Fundamental Laws and Evolution of the Universe), JICFuS, Grants-in-Aid for Scientific Research (JP25400222, JP15H03638, JP17H01101, JP17H01110, JP17H04828, JP18H04337, JP18K03699, and JP19H00697) from the Japan Society for the Promotion of Science (JSPS). 
This research is also supported in part by Interdisciplinary Computational Science Program in Center for Computational Sciences, University of Tsukuba.



\bibliographystyle{mn2e}
\bibliography{tk19,tk19_2}






\bsp	
\label{lastpage}
\end{document}